\documentclass[final,5p]{elsarticle}
\usepackage{graphicx}
\usepackage{dcolumn}% Align table columns on decimal point
\usepackage{amsmath}
\usepackage{amssymb}
\usepackage{amsfonts}
\usepackage{times}
\usepackage{comment}
\usepackage{mathrsfs}
\usepackage{multirow}	
\usepackage{bm}
\usepackage{braket}
\biboptions{numbers,sort&compress}
\usepackage[T1]{fontenc}
\usepackage[utf8]{inputenc}
\usepackage{hyperref}

\usepackage{color}
\usepackage{ulem}

\hypersetup{bookmarksnumbered,%
	colorlinks,%
	linkcolor=blue,%
	citecolor=blue,%
	plainpages=false,%
	pdfstartview=FitH}

\def\fun#1#2{\lower3.6pt\vbox{\baselineskip0pt\lineskip.9pt
\ialign{$\mathsurround=0pt#1\hfil##\hfil$\crcr#2\crcr\sim\crcr}}}

\journal{Phys. Lett. B}

\usepackage[pscoord]{eso-pic}% The zero point of the coordinate systemis the lower left corner of the page (the default).

\newcommand{\placetextbox}[3]{% \placetextbox{<horizontal pos>}{<vertical pos>}{<stuff>}
  \setbox0=\hbox{#3}% Put <stuff> in a box
  \AddToShipoutPictureFG*{% Add <stuff> to current page foreground
    \put(\LenToUnit{#1\paperwidth},\LenToUnit{#2\paperheight}){\vtop{{\null}\makebox[0pt][c]{#3}}}%
  }%
}%

\begin{document}

\begin{frontmatter}

\placetextbox{0.906}{0.96}{
  \normalsize
  NITEP 70
}%
\placetextbox{0.901}{0.945}{
  \normalsize
  May, 2020
}%

\title{Manifestation of the divergence between
antisymmetrized-molecular-dynamics and container pictures of $^{9}$Be via
${}^{9}$Be($p,pn$)${}^{8}$Be knockout reaction}

\author[1,2]{Nguyen Tri Toan Phuc}
\ead{nguyentritoanphuc@yahoo.com} 

\author[3,4]{Mengjiao Lyu\corref{cor1}}
\cortext[cor1]{Corresponding author}
\ead{mengjiao.lyu@nuaa.edu.cn}

\author[6,7]{Yohei Chiba}
\ead{chiba@osaka-cu.ac.jp}

\author[5,6,7]{Kazuyuki Ogata}
\ead{kazuyuki@rcnp.osaka-u.ac.jp}

\address[1]{Department of Nuclear Physics, Faculty of Physics and Engineering
  Physics, University of Science, Ho Chi Minh City, Vietnam}
\address[2]{Vietnam National University, Ho Chi Minh City, Vietnam}
\address[3]{College of Science, Nanjing University of Aeronautics and
Astronautics, Nanjing 210016, China}
\address[4]{Key Laboratory of Aerospace Information Materials and Physics (NUAA), 
MIIT, Nanjing 211106, China}
\address[5]{Research Center for Nuclear Physics (RCNP), Osaka University,
  Ibaraki 567-0047, Japan}
\address[6]{Department of Physics, Osaka City University, Osaka 558-8585,
  Japan}
\address[7]{Nambu Yoichiro Institute of Theoretical and Experimental Physics
  (NITEP), Osaka City University, Osaka 558-8585, Japan}

%\date{\today}

% cSpell: enable

\begin{abstract}
We propose a new approach to probe the spatial extension of the valence neutron orbital in the $^{9}$Be nucleus via the ${}^{9}$Be($p,pn$)${}^{8}$Be knockout reaction. This property of the nuclear molecular orbital has not been
established in previous experimental studies and divergence exists
between the theoretical descriptions of ${}^{9}$Be from different
perspectives, \textit{i.e.}, the antisymmetrized molecular dynamics and the
container pictures of cluster dynamics. These pictures are represented by two different well-proven microscopic models, the antisymmetrized molecular dynamics (AMD) and Tohsaki-Horiuchi-Schuck-R\"{o}pke (THSR) wave functions. The corresponding reduced width amplitudes (RWAs) in the $^{8}$Be$+n$ channel are extracted from both the AMD and THSR wave functions, and they are found to describe drastically different
valence-nucleon motion, which shows the theoretical ambiguity in describing
the  $\pi$-orbitals in $^{9}$Be. Using the RWAs as input, the physical
observables of the ${}^{9}$Be($p,pn$)${}^{8}$Be knockout reaction are
predicted by the distorted-wave impulse approximation (DWIA) framework. The
magnitudes of the triple-differential cross sections (TDX) are found to be
highly sensitive to the RWA input. It is concluded that the
${}^{9}$Be($p,pn$)${}^{8}$Be knockout reaction could provide a feasible
probing for the subtle differences between several structure models manifesting through the spatial extension of the $\pi$-orbital in the $^{9}$Be
nucleus.	
\end{abstract}


\end{frontmatter}

\section{Introduction}
\label{sec:intro}

In atomic nuclei, the dynamical assembling of nucleons into clusters is a
fundamental aspect of nuclear many-body dynamics. Most notably, the
formation of $\alpha$-clusters that saturates the spin-isospin degrees of
freedom provides a strong binding to the nucleus \cite{Hor12,Fre18}. For
neutron-rich nuclei, the coupling between the $\alpha$-clusters and the valence neutrons gives rise to the covalent molecular bindings, which constitute the
``nuclear molecules'' \cite{Oer06}. The Be isotopes are the most prominent
examples of nuclear molecular states, where valence neutrons occupy the molecular orbitals around the 2$\alpha$ core, as predicted by several theoretical studies
using a wide variety of microscopic models \cite{Oer06,
Fre18, Kim16, Eny12, Eny03, Eny18}. In this study, we focus on ${}^{9}$Be, which provides the most straightforward interpretation of the nuclear molecular orbital concept. The molecular structure of the Borromean ${}^{9}$Be also plays a major role in the explosive nucleosynthesis scenario through the $\alpha(\alpha n,\gamma){}^{9}$Be$(\alpha,n){}^{12}$C r-process \cite{Sas05}, which provides an alternative path to the triple-alpha reaction for the production of ${}^{12}$C. Therefore, an accurate description of $^{9}$Be molecular states has long been required in the calculations of $\alpha(\alpha n,\gamma){}^{9}$Be reaction and its inverse process ${}^{9}$Be$(\gamma,n)$ \cite{Cas14,Ods19}.

In the calculations using the generator coordinate method (GCM) in
Ref.~\cite{Oka77}, the molecular orbitals of $\pi$-character and
$\sigma$-character are constructed from the linear combination of ``atomic''
orbitals (LCAO) in $p$-wave around each of the two $\alpha$-clusters in the
$^{9,10}$Be isotopes. In later studies \cite{Oka79, Ita00}, linear
combinations of Gaussian wave packets are widely adopted in various molecular
orbit models. The ansatz of molecular orbitals is confirmed by the studies of
antisymmetrized molecular dynamics (AMD), which reproduces the molecular
states of Be isotopes without any model assumption \cite{Eny95}. In Ref.~\cite{Ita00}, the spatial extension of the $\pi$-orbitals are
found to be critical in reproducing the energy difference between the
$3/2^{-}$ and the $1/2^{-}$ states of $^{9}$Be, and the valence neutrons are
predicted to locate at a much more extended position measured from the center of
$\alpha$-cluster, as compared to the compact orbitals constructed by LCAO. In
the generalized two-center cluster model (GTCM) \cite{Ito04}, the transition
from the molecular orbitals to the di-cluster states are demonstrated and the
prevalence of molecular configurations are confirmed for $^{10}$Be when the
$\alpha$-$\alpha$ distance $S<4$ fm. In Refs.~\cite{Lyu15, Lyu16}, the
$\pi$-orbitals are reproduced by the variational calculations of $^{9,10}$Be
isotopes using the Tohsaki-Horiuchi-Schuck-R\"{o}pke (THSR) wave functions. These THSR wave functions, originally formulated to describe the gaslike structure of $\alpha$-conjugate systems \cite{Toh01}, indicate a nonlocalized motion for the $\alpha$-clusters under the external covalent binding from the valence neutrons. The THSR calculation \cite{Li20} also shows that the $\pi$-orbital neutrons can affect the lower region of $^{9,10}$Be momentum distributions.  

In previous theoretical studies, special attention has been paid to the essential
role of nuclear molecular structures in reproducing various properties of Be
isotopes, especially the spectral structure of rotational bands \cite{Fre18,Oer06}.
On the other hand, the spatial information on the molecular orbitals is rarely discussed except for a few theoretical calculations such as in Refs.~\cite{Oka79,Eny03,Lyu16,Ara03}. In
Ref.~\cite{Oka79}, the nuclear $\pi$-orbit described by the GCM wave function
of $^{9}$Be is discussed via a ``two-dimensional reduced width amplitude.'' In
Ref.~\cite{Eny03}, the sketches of molecular orbits are extracted from the AMD
wave functions of $^{10}$Be. In Refs.~\cite{Lyu15,Lyu16}, the molecular orbits
are discussed through the intrinsic densities calculated from the THSR wave
functions of $^{9}$Be and $^{10}$Be. However, the spatial extension of the
valence neutron and its model dependence have not been confirmed yet by the
experimental observables. We remark that although electron scattering data are often used to benchmark the wave function of several models \cite{Oka77,Oka79}, they only probe the bulk property of the nucleus, i.e., the one-body diagonal and transition densities, and do not provide any direct information about the valence nucleon motion.  

Essentially, both the AMD and the THSR wave functions are variational
approaches taking into account the antisymmetrization effects. It was shown
that both the AMD and THSR can reasonably describe the known properties of the
ground and excited states of ${}^{9}$Be, such as the radius, energies, and
electromagnetic transitions \cite{Eny95, Lyu15}. Meanwhile, the different
physical perspectives between these two microscopic models induce divergence
in the descriptions of $^{9}$Be. The AMD wave function focuses on the
independent motion of single particles, as a common nature of the molecular
dynamics models, and the $\alpha$-clusters and molecular orbitals are produced
with no \textit{a priori} assumption on the structural information
\cite{Eny03,Eny12,Eny18,Kim16}. On the other hand, in the container picture \cite{Zho14,Zho20} the THSR wave function is built upon the Brink-Bloch type of bases \cite{Bri66}, which emphasizes the nonlocalized motion of the nuclear clusters formulated with strong internal correlations, and the cluster dynamics is constrained within a parameterized Gaussian container, which is determined by variational calculations \cite{Lyu15}. It is known that the AMD and THSR wave functions emphasize two different types of duality. While the cluster and shell-model structures can coexist in AMD formalism \cite{Eny18}, the container picture of THSR wave function can flexibly describe the nonlocalized gaslike and localized crystal-like states of nucleus \cite{Zho20}. The divergence between the AMD and THSR pictures is significant especially in the description of valence neutron motion, as it is more sensitive to the binding mechanism of the nucleus. 

Recently, the development of experimental facilities has enabled new
possibilities to clarify the single-nucleon properties in nuclei by using the
proton-induced nucleon knockout reactions in normal \cite{Wak17} and inverse
kinematics \cite{Phu19}. Given the appropriate kinematical condition, these ($p,pN$) knockout reactions provide the deepest probe for the single-particle wave function among the hadronic-induced reactions \cite{Aum13}. They can also probe the neutron orbital in a large variety of nuclei, including unstable ones, which are currently not feasible for electron-induced $(e,e'p)$ reactions. The theoretical descriptions of the ($p,pN$) knockout reactions have been well established in the distorted-wave impulse approximation (DWIA) framework \cite{Wak17}, which has been successfully applied in the analyses of experimental results for a wide range of
nuclei \cite{Phu19, Wak17, Kaw18, Che19, Tan19, Tan20, Yan21}.
In addition, the cross sections of the ($p,p\alpha$) reactions on the
$^{10,12}$Be targets has been predicted for the probing of $\alpha$-clustering
effects \cite{Lyu18, Lyu19} in the same DWIA framework \cite{Yos16, Yos18-2}. The
reliability of this framework has been convincingly demonstrated in the
parameter-free ($p,p\alpha$) calculation for $^{20}$Ne \cite{Yos19} and Sn \cite{Tan21} targets.  

In this study, we investigate the divergence between the molecular dynamics
and container pictures of $^{9}$Be nucleus, in the AMD and THSR models
respectively, through the structural calculations of model wave functions and
the predictions of reaction observables of proton-induced nucleon knockout
reactions. A new approach is proposed to probe the spatial extension of the
$\pi$-orbital in the ground state of the $^{9}$Be target via the
$^{9}$Be$(p,pn)^{8}$Be knockout reaction, with the triple-differential cross
sections (TDX) predicted in the DWIA framework incorporated with the microscopic reduced width amplitude (RWA). We show that the magnitudes of the TDX curves are strongly sensitive to the different ($\pi$-orbital) neutron distributions of the $^{9}$Be target described by the AMD and THSR wave functions, which demonstrate the potential of this approach as a direct probe for the spatial extensions of molecular orbitals.

This article is organized as follows: In Sec.~\ref{sec:dwia}, the DWIA
framework for the ($p,pn$) knockout reaction is introduced. In
Sec.~\ref{sec:wf}, the AMD and THSR wave functions of $^{9}$Be, and the
extraction of corresponding RWA in the $^{8}$Be$+n$
channel are explained. Numerical inputs for the ($p,pn$) knockout reaction and
the TDX results with different distorting potentials are introduced and
discussed in Sec.~\ref{sec:results}. The last Sec.~\ref{sec:sum} contains the
conclusion.

\section{DWIA framework}
\label{sec:dwia}

The ${}^{9}$Be($p,pn$)${}^{8}$Be knockout reaction is analyzed with the same
DWIA framework as presented in Refs.~\cite{Wak17,Yos18}. The transition amplitude
for the ($p,pn$) reaction is given by
\begin{align}
T_{\bm{K}_0\bm{K}_1\bm{K}_2}
&=
\Braket{
	\chi_{1,\bm{K}_1}^{(-)}\chi_{2,\bm{K}_2}^{(-)}
	|t_{pn}|
	\chi_{0,\bm{K}_0}^{(+)}\varphi_{n}
},
\label{eq:trans}
\end{align}
where $\chi_{i,\bm{K}_i} (i=0,1,2)$ are the distorted scattering wave
functions of the $p$-${}^{9}$Be, $p$-${}^{8}$Be, and $n$-${}^{8}$Be systems,
respectively. The outgoing and incoming boundary conditions of these
scattering waves are specified by the superscripts $(+)$ and $(-)$,
respectively. $\bm{K}_i$ is the momentum of particle $i$ in the three-body
center-of-mass (c.m.) frame. The relative wave function of the valence neutron
bound inside ${}^{9}$Be is denoted as $\varphi_{n}$, whose radial part is the
RWA as defined in Ref.~\cite{Hor77}. $t_{pn}$ is the transition operator for
the $p$-$n$ scattering. The TDX in the laboratory frame of the
${}^{9}$Be($p,pn$)${}^{8}$Be reaction is given by
\begin{align}\label{eq:tdx}
\frac{d^3 \sigma}{dE_1^{\rm L} d\Omega_1^{\rm L} d\Omega_2^{\rm L}}
=&F_{\rm kin}C_0
\frac{d\sigma_{pn}}{d\Omega_{pn}}
\left| \bar{T}_{{\bm K}_0{\bm K}_1{\bm K}_2} \right| ^2, \\
\bar{T}_{{\bm K}_0{\bm K}_1{\bm K}_2}
=&\int d\bm{R}\,
\chi_{1,{\bm K}_1}^{(-)}({\bm R})
\chi_{2,{\bm K}_2}^{(-)} ({\bm R})\nonumber \\ 
&\quad\times
\chi_{0,{\bm K}_0}^{(+)}({\bm R})\varphi_{n}(\bm{R})
e^{-i\bm{K}_0\cdot\bm{R} / A},
\end{align}
where $F_{\rm kin}$ and $C_0$ are kinematical factors, and
$d\sigma_{pn}/d\Omega_{pn}$ is the $p$-$n$ differential cross section at the
energy and the scattering angle deduced from the ($p,pn$) kinematics. The
M\o ller factor is taken into account when $d\sigma_{pn}/d\Omega_{pn}$ is
evaluated in Eq.~(\ref{eq:tdx}). In complement to the TDX, the transition amplitude in Eq.~(\ref{eq:trans}) can also be used to calculate the longitudinal and transverse momentum distributions \cite{Phu19,Yos18}, which are the usually measured quantities in inverse kinematics ($p,pN$) experiments with radioactive beams.

\section{AMD and THSR wave functions}
\label{sec:wf}
In this study, we use two well-established models, the AMD and THSR wave
functions, for the description of the $^{9}$Be target. Here, we introduce
briefly the formulations of both wave functions. The detailed explanations can
be found in Refs.~\cite{Chi15} and \cite{Lyu15} for the AMD and the THSR wave
functions, respectively.

In AMD, basis wave functions of the system are given in the Slater
determinants form 
\begin{equation}\label{eq:amd}
  \Phi_{\textrm{AMD}}=\mathcal{A}\left\{\psi_{1} \psi_{2} \ldots \psi_{A}\right\},
\end{equation}
where $\psi_{i}=\phi_{i}\otimes\chi_{i}\otimes\xi_{i}$ is the nucleon wave
function, which contains a deformed Gaussian term in the coordinate space
\begin{equation}
\phi_{i}\propto
  \exp \left[
    -\sum_{\sigma=x y z} 
      \nu_{\sigma}\left(
        r_{i \sigma}
        -\frac{Z_{i \sigma}}{\sqrt{\nu_{\sigma}}}
      \right)^{2}\right].
\end{equation}
Here, coordinates $\bm{Z}_{i}$ are the centroids of the deformed Gaussians
satisfying condition $\sum_{i=1}^{A}\bm{Z}_{i}=\bm{0}$. The $\chi_{i}$ and
$\xi_{i}$ are the spin and isospin components, respectively, and they are
defined as
\begin{equation}
\chi_{i}=
  \alpha_{i}\ket{\uparrow}+\beta_{i}\ket{\downarrow}, 
  \quad \xi_{i}=\ket{p} \textrm { or }\ket{n}.
\end{equation}
The basis wave functions are prepared by the energy variation of the AMD wave
functions in Eq.~(\ref{eq:amd}) after parity projection, where constraint on
the expectation values of harmonic oscillator quanta are imposed. The total
wave function of the $^{9}$Be target is obtained by diagonalizing the
Hamiltonian with respect to the projected AMD bases $\Phi_{\textrm{AMD}}^{JM}$
with total spin $J$ and $z$-component $M$ after angular momentum projection
\cite{Hor86}.

In THSR, basis wave functions of the system are constructed in the Brink-Bloch
form for the two $\alpha$ clusters and the valence neutron, as
\begin{equation}
\begin{aligned}
&\Phi_{\textrm{Brink}}(
  \bm{R}_{1},\bm{R}_{2}, \bm{R}_{n} )\\
&\quad=\mathcal{A}\{
  \psi_{1}(\bm{r}_{1},\bm{R}_{1})
  \psi_{2}(\bm{r}_{2},\bm{R}_{1})
  \dots
  \psi_{4}(\bm{r}_{2},\bm{R}_{1})\\
&\phantom{\quad=\mathcal{A}\{} \times  
  \psi_{5}(\bm{r}_{5},\bm{R}_{2})
  \psi_{6}(\bm{r}_{6},\bm{R}_{2})
  \dots
  \psi_{8}(\bm{r}_{8},\bm{R}_{2})\\
&\phantom{\quad=\mathcal{A}\{} \times  
  \psi_{9}(\bm{r}_{9},\bm{R}_{n})
\},
\end{aligned}
\end{equation}
where $\bm{R}_{1,2}$ and $\bm{R}_{n}$ are generator coordinates of two $\alpha$
clusters and the valence neutron, respectively. The
$\psi_{i}=\phi_{i}\otimes\chi_{i}\otimes\xi_{i}$ is the nucleon wave function,
which is the product of a Gaussian term in the coordinate space
\begin{equation}
\phi_{i}(\bm{r}_{i},\bm{R})\propto
  \exp \left[
    -\nu( \bm{r}_{i} -\bm{R} )^{2}
  \right],
\end{equation}
and the spin-isospin terms
\begin{equation}
  \chi_{i}= \ket{\uparrow} \textrm { or }\ket{\downarrow}, 
  \quad \xi_{i}=\ket{p} \textrm { or }\ket{n}.
\end{equation}
The THSR wave function of the $^{9}$Be target is then formulated as
\begin{equation}\label{eq:thsr}
\begin{aligned}
&\Phi_{\textrm{THSR}}=  \int d\bm{R}
  \,\mathcal{G}(\bm{R},\bm{\beta})
  \int d\bm{R}_{n}
  \,\mathcal{G}(\bm{R}_{n},\bm{\beta}_{n})
  e^{i\phi\left(\bm{R}_{n}\right)}\\
&\phantom{\Phi_{\textrm{THSR}}=  \int}
\times\Phi_{\textrm{Brink}}(
    \bm{R},-\bm{R}, \bm{R}_{n}
),
\end{aligned}
\end{equation}
where $\mathcal{G}(\bm{R},\bm{\beta})$ is the deformed Gaussian function
\begin{equation}
\mathcal{G}(\bm{R}, \bm{\beta})
=\exp \left(
  -\frac{R_{x}^{2}+R_{y}^{2}}{\beta_{x y}^{2}}
  -\frac{R_{z}^{2}}{\beta_{z}^{2}}
\right),
\end{equation}
and $\phi(\bm{R}_{n})$ is the azimuthal angle in the spherical coordinates of
$\bm{R}_{n}$ to produce the necessary negative parity of the $\pi$-orbital in
the $^{9}$Be target. The gaslike and non-gaslike natures of a state are characterized by the size parameters $\bm{\beta}$ and $\bm{\beta}_{n}$ of the Gaussian containers. The center-of-mass projection and the angular momentum projection are imposed on the THSR wave function to restore the translational and the rotational symmetries of $^{9}$Be \cite{Lyu15}.

The structure information of $^{9}$Be is integrated into the DWIA model
through the RWA in the $^{8}$Be$+n$ channel, extracted from the AMD and the
THSR wave functions of the $^{9}$Be target as
\begin{equation}
\begin{aligned}
&y(a) = \\
&\quad\sqrt{9}
\left. \left\langle
  \frac{\delta(r-a)}{r^{2}} 
  \phi(^{8}\textrm{Be}) 
  [
    Y_{1}(\hat{r}) 
    \chi_{n}
  ]
  \right|\Phi(^{9}\textrm{Be}) 
\right\rangle,
\end{aligned}
\end{equation}
where $\phi(^{8}\textrm{Be})$ is the wave function of the $^{8}$Be residual in
the first $0^{+}$ resonance state, $\chi_{n}$ is the spin $1/2$ of the valence
neutron, and $\Phi(^{9}\textrm{Be})$ is the AMD or THSR wave function of the
$^{9}$Be target in the $3/2^{-}$ ground state. In our calculation,
$\phi(^{8}\textrm{Be})$ is obtained by solving the $\alpha$-$\alpha$
scattering problem with the Ali-Bodmer potential \cite{Ali66}, which is fitted
from the experimental phase shift. This choice provides a realistic wave
function for $^{8}$Be. The RWA using AMD wave function is calculated with the
Laplace expansion method proposed in Ref.~\cite{Chi17}, and the RWA using the
THSR wave function is calculated with a new method \cite{Lyu20}, which is an
extended version of the traditional Brink expansion method used in
Ref.~\cite{Eny03b}.

\section{Results}
\label{sec:results}

We investigate the $^{9}$Be$(p,pn)$$^{8}$Be reactions at 392 MeV in the
laboratory frame with the following kinematical conditions. The kinetic
energies for the incoming and outgoing protons are taken at 392 and 251 MeV,
respectively. The emission angle of the outgoing proton is chosen to be
($\theta_1^L, \phi_1^L$) = ($34^\circ, 0^\circ$). The angle $\theta_2^L$ of
the emitted neutron is varied in the range of $17^\circ$--$84^\circ$ while the
angle $\phi_2^L$ is fixed at $180^\circ$. The quasifree (recoilless) condition
is found to be fulfilled at around $\theta_2^L=50.5^\circ$. The Franey-Love
parametrization \cite{FL85} is used for the $p$-$n$ scattering transition
interaction $t_{pn}$. The relativistic kinematics is adopted for all
scattering particles. 

\begin{figure}[htbp]
	\centering
	\includegraphics[width=0.48\textwidth]{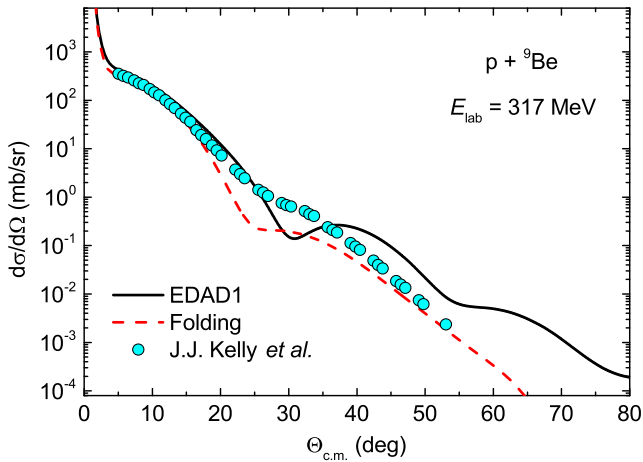}
	\caption{Differential cross section of $p$ + $^{9}$Be elastic scattering at
		317 MeV. The cross sections calculated with EDAD1 (black solid line) and
		folding potentials (red dashed line) are compared with the experimental
		data taken from Ref.~\cite{Kel91}.}
	\label{fig.elasticXsec}
\end{figure}

The distorting potentials, which generate the scattering wave functions in Eq.~(\ref{eq:trans}), are the crucial inputs in the DWIA calculation. To
estimate the uncertainty coming from the optical potential, we perform the TDX
calculations with two sets of optical potentials. First, the EDAD1
parametrization of the Dirac-phenomenology potential \cite{Coo93} is used. The
other is the microscopic single-folding model \cite{Min10} using the Melbourne
\textit{g}-matrix \cite{Amos00} and densities provided by the same wave
functions used in the RWA calculations. As shown in
Fig.~\ref{fig.elasticXsec}, both choices of optical potential reasonably
reproduce the $\Theta_\text{c.m.}<20^\circ$ region of $p$ + $^{9}$Be elastic scattering experimental data at 317 MeV \cite{Kel91} without using additional parameters. Although the results of the EDAD1 (folding) potential slightly overestimates (underestimates) the data in the $\Theta_\text{c.m.}>20^\circ$ region, the contributions to the TDX from the wave functions corresponding to this region is rather small (note the cross section scale). The minor differences around the data in the results from the two types of potentials justify their use to estimate the uncertainty in the DWIA calculation with the chosen kinematical condition. Comparison with the same experimental data also suggests that the EDAD1 set is the most suitable version among various Dirac-phenomenology potentials \cite{Coo93,Coo09} for the considered reaction. We note that although the spin-orbit component of the optical potential is required to
reproduce the data in Fig.~\ref{fig.elasticXsec}, its effect is very small in
the TDX calculation of the considered kinematics \cite{Nor19}. The nonlocality
correction, which is essential for a quantitative ($p,pn$) description
\cite{Phu19,Wak17}, is taken into account by multiplying the distorted wave
functions by the Darwin (for Dirac potential) or Perey (for folding
potential) factors with the range parameter $\beta_\text{NL}=0.85$ fm. 

\begin{figure}[htbp]
	\centering
	\includegraphics[width=0.48\textwidth]{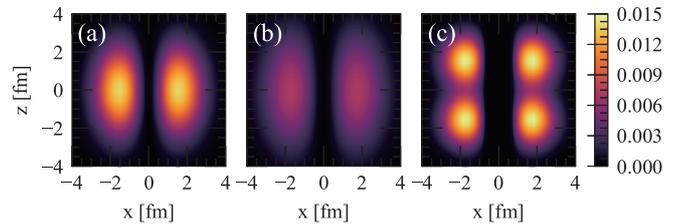}
  \caption{Comparison of the intrinsic density distributions $\rho(\bm{r})$ of
  the valence neutron in the ground state of $^{9}$Be, predicted by the AMD
  (a), THSR (b), and a simple molecular orbital (c) wave functions. The unit of the density plotting is
  fm$^{-3}$. $\rho(\bm{r})$ is normalized as $\int d\bm{r}\, \rho(\bm{r})=1$.
  }
	\label{fig.mo}
\end{figure}

In the AMD calculations, the Gogny D1S \cite{Ber91} interaction is adopted. The THSR calculations use the Volkov No.~2 \cite{Vol65} and G3RS \cite{Yam79} as the central and spin-orbit interactions, respectively. The physical properties of $^{9}$Be calculated in both approaches along with experimental values \cite{Til04} are listed in Table \ref{tab:9be}. It shows that both wave functions can well describe the energy and radius of the $3/2^{-}$ ground state. 

%%%%%%%%%%%%%%%%%%%%%%%%%%%%%%
\begin{table}[h]
	\begin{center}
		\caption{Physical properties of $^{9}$Be calculated in the AMD and THSR
			approaches. $E$ denotes the energy of the $3/2^{-}$ ground state and $\Delta
			E$ is the excitation energy of the first $5/2^{-}$ state. $R_{\textrm{rms}}$
			is the matter root-mean-square radius of the $3/2^{-}$ ground state. ``Exp.'' denotes
			corresponding experimental values \cite{Til04}.}
		\label{tab:9be}
		\begin{tabular}{cccc}
			\hline
			&$E$ (MeV) &$\Delta E$ (MeV)  &$R_{\textrm{rms}}$ (fm)\\
			\hline
			AMD   &-59.2    &1.8    &2.52    \\
			THSR  &-55.4    &2.4    &2.55    \\
			Exp.  &-58.2    &2.4    &2.45    \\
			\hline
		\end{tabular}
	\end{center}
\end{table}
%%%%%%%%%%%%%%%%%%%%%%%%%%%%%%

\begin{figure}[htbp!]
	\centering
	\includegraphics[width=0.48\textwidth]{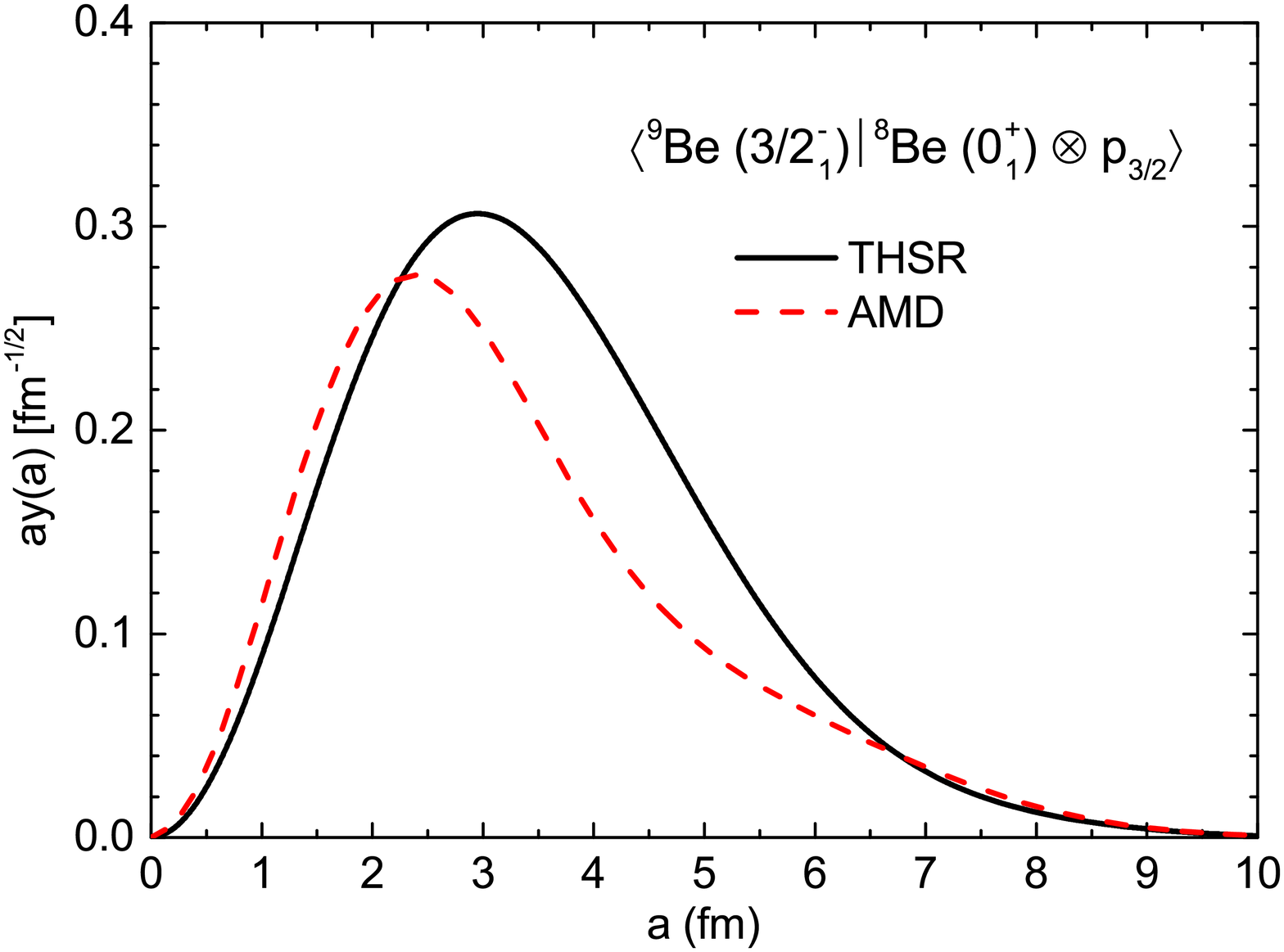}
	\caption{The RWAs of $^{9}\text{Be}(3/2^-_1)$ state in
		$^{8}\text{Be}(0^+_1)+n$ configuration given by AMD and THSR
		calculations.}
	\label{fig.9Be-RWA}
\end{figure}

In Fig.~\ref{fig.mo}, we compare the intrinsic density distributions of the
valence neutron occupying the $\pi$-orbit in the ground state of $^{9}$Be,
predicted by the AMD, THSR, and simple molecular orbital wave functions.
Both the AMD and THSR models naturally display the $\pi$-orbital pattern
without any assumption of the molecular orbital structure. The valence
neutron in the AMD and THSR models is more freely distributed, in contrast
to the simple molecular orbital configuration, where it is mostly localized
around the two $\alpha$ clusters. Here, it is clearly shown that the AMD
wave function describes a relatively more compact density distribution of
the valence neutron, while the THSR wave function corresponds to a much
larger spatial extension in the horizontal direction, which is related to
the neutron nonlocalized motion in its container.

In Fig.~\ref{fig.9Be-RWA}, we illustrate the RWAs calculated from the THSR
(black solid line) and AMD (red dashed line) wave functions. The corresponding
neutron spectroscopic factors are 0.259 and 0.182, respectively. It shows that
the RWA of AMD basis is more compact compared to the one from the THSR basis. This
implies a clear difference in the valence neutron motion between the AMD and
THSR wave functions. The suppression in the surface region of the AMD-RWA is
caused by the cluster breaking component included in the AMD model space,
which is one of the fundamental differences between the molecular dynamics of
independent nucleons and the container model of cluster dynamics.  It would be
interesting to perform the in-depth analyses for the RWA by structural
investigations, but it is beyond the scope of the present work. Instead, the
ambiguities in the theoretical predictions of RWAs presented in
Fig.~\ref{fig.9Be-RWA} can be clearly clarified via the experimental
observables of the $^{9}$Be$(p,pn)^{8}$Be reaction. Finally, we note that the spectroscopic factor values in our calculations is somewhat smaller than those observed in transfer reactions analyses (see, e.g., Ref.~\cite{Tsa05}). This discrepancy can be solved by using a bound-state approximation for $^{8}$Be wave function as in Ref.~\cite{Oka79}. Since this approximation affects both THSR and AMD results with the same magnitude, whether or not we use it does not affect our discussion below.

\begin{figure}[htbp!]
	\centering
	\includegraphics[width=0.48\textwidth]{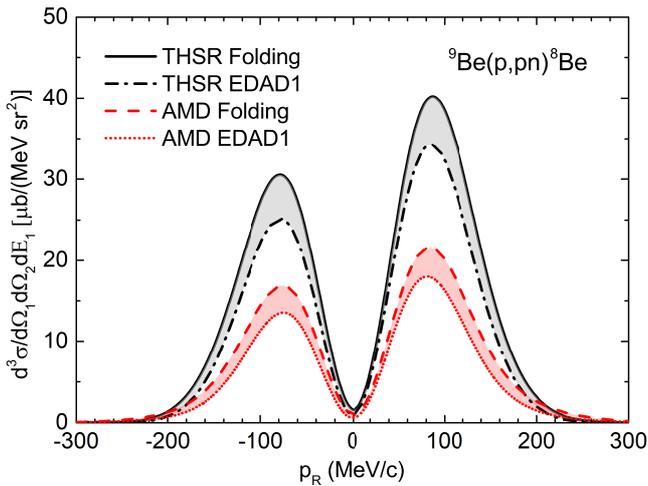}
	\caption{The TDXs of $^{9}\text{Be}$($p$,$pn$)$^{8}\text{Be}(0^+_1)$ reaction at
		392 MeV. $p_R$ is the recoiled momentum.}
	\label{fig.TDX-9Be}
\end{figure}

In Fig.~\ref{fig.TDX-9Be}, the TDX of $^{9}\text{Be}$($p$,$pn$)$^{8}\text{Be}(0^+_1)$ reaction given by the DWIA calculations are compared
for the RWA from AMD (black lines) and THSR (red lines) wave functions. For
each RWA, we use two sets of distorting potentials as described above. The TDX
with THSR wave function is larger than the one by AMD by nearly a factor of 2.
The shaded area represents the uncertainty of the TDX coming from that of the
optical potentials. Figure \ref{fig.TDX-9Be} illustrates that the TDX from the
two microscopic models are well separated, especially around the two peaks.
This drastic difference in the magnitude of the TDX between the two justifies
the use of $(p,pn)$ as a sensitive probe for the molecular orbit of neutron in
$^{9}$Be. By comparing the theoretical predictions in Fig.~\ref{fig.TDX-9Be}
with experimental data, differentiation between the antisymmetrized molecular
dynamics and the container picture will become feasible for the description of
the ground state of $^{9}$Be. We also perform similar calculations for the $^{9}\text{Be}$($p$,$pn$)$^{8}\text{Be}$ longitudinal and transverse momentum distributions and obtain the same difference in the magnitudes corresponding to the two structure models. These results suggest that the approach presented in this study can be applied to more exotic nuclei by analyzing their $(p,pn)$ reactions in inverse kinematics experiments.

\section{Conclusion and outlooks} \label{sec:sum}
In this work, we have investigated the divergence between the antisymmetrized
molecular dynamics and the container pictures in the theoretical descriptions of the $^{9}$Be nucleus and proposed to probe the corresponding spatial extensions of the nuclear $\pi$-orbit in $^{9}$Be via the ${}^{9}$Be($p,pn$)${}^{8}$Be knockout reaction. The $^{9}$Be target in the initial state is described by using two different well-established microscopic models including the AMD and THSR wave functions. The RWAs in the $^{8}$Be$+n$ channel are extracted from these two structural models, which are found to describe the drastically different motion of the valence neutron. The clear difference between these two RWAs demonstrates the theoretical ambiguity in describing the $\pi$-orbital of $^{9}$Be nucleus. 

Using the RWAs as input, the physical observables of the ${}^{9}$Be($p,pn$)${}^{8}$Be
knockout reaction are predicted in the DWIA framework. The magnitudes of the
TDX curves are found to be highly sensitive to the different spatial extensions
of the valence nucleon motion. It is concluded that the
${}^{9}$Be($p,pn$)${}^{8}$Be knockout reaction could provide a feasible
probing for the size of the $\pi$-orbital in the $^{9}$Be nucleus. By incorporating the wave functions into this framework, other structure models such as \textit{ab initio} \cite{Pie02,For05} and cluster shell model \cite{Del18} can also be benchmarked. The present approach can be naturally extended for more exotic nuclei using inverse kinematics $(p,pN)$ reactions with radioactive beams. It is expected that in combination with traditional observables, the experimental results for the reaction proposed in this study will provide better validation and constraint for the development of nuclear structure models.	

% cSpell: disable

\section*{Acknowledgments}
The computation was carried out with the computer facilities at the Research
Center for Nuclear Physics, Osaka University. One of the authors (N.T.T.P)
thanks the Research Center for Nuclear Physics, Osaka University
for hospitality during his stay in which this work was initiated. He also
wants to thank the Vietnam MOST for its support through the Physics
Development Program Grant No. {\DJ}T{\DJ}LCN.25/18. This work was supported in
part by Grants-in-Aid of the Japan Society for the Promotion of Science
(Grants No. JP16K05352).

%%--------------------------------------------------------------------%%
%%                           References                               %%
%%--------------------------------------------------------------------%%


\begin{thebibliography}{00}
% nuclear molecule, extended ikeda diagram
\bibitem{Hor12}H. Horiuchi, K. Ikeda, K. Kat{\=o}, 
\href{https://doi.org/10.1143/PTPS.192.1}{Prog. Theor. Phys. 192 (2012) 1–238}.
\bibitem{Fre18}M. Freer, H. Horiuchi, Y. Kanada-En'yo, D. Lee, Ulf-G. Mei{\ss}ner,
\href{https://doi.org/10.1103/RevModPhys.90.035004}{Rev. Mod. Phys. 90 (2018) 035004}.
\bibitem{Oer06}W. von Oertzen, M. Freer, Y. Kanada-En'yo,
\href{https://doi.org/10.1016/j.physrep.2006.07.001}{Phys. Rep. 432 (2006) 43–113}.

% additonal reviews for nuclear molecules
\bibitem{Eny18}Y. Kanada-En’yo, H. Horiuchi, 
\href{https://doi.org/10.1007/s11467-018-0830-y}{Front. Phys. 13 (2018) 132108}.
\bibitem{Kim16}M. Kimura, T. Suhara, Y. Kanada-En'yo,
\href{https://doi.org/10.1140/epja/i2016-16373-9}{Eur. Phys. J. A 52 (2016) 373}.
\bibitem{Eny12}Y. Kanada-En'yo, M. Kimura, A. Ono,
\href{https://doi.org/10.1093/ptep/pts001}{Prog. Theor. Exp. Phys. 2012 (2012) 1A202-0}.
\bibitem{Eny03}Y. Kanada-En’yo, M. Kimura, H. Horiuchi, 
\href{https://doi.org/10.1016/S1631-0705(03)00062-8}{C. R. Phys. 4 (2003) 497–520}.

% 8Be(n,y)9Be paper
\bibitem{Sas05}T. Sasaqui, K. T. Kajino, G. Mathews, K. Otsuki, T. Nakamura, 
\href{https://doi.org/10.1086/497061}{Astrophys. J. 634 (2005) 1173–1189}.
\bibitem{Cas14}J. Casal, M. Rodr\'{i}guez-Gallardo, J.M. Arias, I.J. Thompson,
\href{https://doi.org/10.1103/PhysRevC.90.044304}{Phys. Rev. C 90 (2014) 044304}.
\bibitem{Ods19}M. Odsuren, Y. Kikuchi, T. Myo, H. Masui, K. Kat{\=o}, 
\href{https://doi.org/10.1103/PhysRevC.99.034312}{Phys. Rev. C 99 (2019) 034312}.


% be isotopes, GCM
\bibitem{Oka77}S. Okabe, Y. Abe, H. Tanaka,
\href{https://doi.org/10.1143/PTP.57.866}{Prog. Theor. Phys. 57 (1977) 866–881}.
\bibitem{Oka79}S. Okabe, Y. Abe,
\href{https://doi.org/10.1143/PTP.61.1049}{Prog. Theor. Phys. 61 (1979) 1049–1064}.
% \bibitem{Sey81}M. Seya, M. Kohno, S. Nagata,
% \href{https://doi.org/10.1143/PTP.65.204}{Prog. Theor. Phys. 65 (1981) 204–223}.


% 9be,10be, MO+GCM, the flaw in LCAO
\bibitem{Ita00}N. Itagaki, S. Okabe,
\href{https://doi.org/10.1103/PhysRevC.61.044306}{Phys. Rev. C 61 (2000) 044306}.


% be isotopes, AMD
\bibitem{Eny95}Y. Kanada-En'yo, H. Horiuchi, A. Ono,
\href{https://doi.org/10.1103/PhysRevC.52.628}{Phys. Rev. C 52 (1995) 628–646}.
%\bibitem{Eny99}Y. Kanada-En'yo, H. Horiuchi, A. Dot{\'e},
%\href{https://doi.org/10.1103/PhysRevC.60.064304}{Phys. Rev. C 60 (1999) 064304}.

% 10Be, GTCM, evolution of MO to di-cluster
\bibitem{Ito04}M. Ito, K. Kat{\=o}, K. Ikeda, \href{https://doi.org/10.1016/j.physletb.2004.01.090}{Phys. Lett. B 588 (2004) 43–48}. 

% 9be,10be, THSR
\bibitem{Lyu15}M. Lyu, Z. Ren, B. Zhou, Y. Funaki, H. Horiuchi, G. R\"{o}pke, P. Schuck, A. Tohsaki, C. Xu, T. Yamada,
\href{https://doi.org/10.1103/PhysRevC.91.014313}{Phys. Rev. C 91 (2015) 014313}.
\bibitem{Lyu16}M. Lyu, Z. Ren, B. Zhou, Y. Funaki, H. Horiuchi, G. R\"{o}pke, P. Schuck, A. Tohsaki, C. Xu, T. Yamada,
\href{https://doi.org/10.1103/PhysRevC.93.054308}{Phys. Rev. C 93 (2016) 054308}.
\bibitem{Toh01}A. Tohsaki, H. Horiuchi, P. Schuck, G. R\"{o}pke,
\href{https://doi.org/10.1103/PhysRevLett.87.192501}{Phys. Rev. Lett. 87 (2001) 192501}.
\bibitem{Li20}S. Li, T. Myo, Q. Zhao, H. Toki, H. Horiuchi, C. Xu, J. Liu, M. Lyu, Z. Ren,
\href{https://doi.org/10.1103/PhysRevC.101.064307}{Phys. Rev. C 101 (2020) 064307}.

% 9Be, spatial information, RGM, Arai
\bibitem{Ara03}K. Arai, P. Descouvemont, D. Baye, W.N. Catford,
\href{https://doi.org/10.1103/PhysRevC.68.014310}{Phys. Rev. C 68 (2003) 014310}.

% be isotopes, GCM, descouvemount
%\bibitem{Des02}P. Descouvemont,
%\href{https://doi.org/10.1016/S0375-9474(01)01286-6}{Nucl. Phys. A699 (2002) 463–478}.

% Container picture
\bibitem{Zho14}B. Zhou, Y. Funaki, H. Horiuchi, Z. Ren, G. R\"{o}pke, P. Schuck, A. Tohsaki, C. Xu, T. Yamada,
\href{https://doi.org/10.1103/PhysRevC.89.034319}{Phys. Rev. C 89 (2014) 034319}.
\bibitem{Zho20}B. Zhou, Y. Funaki, H. Horiuchi, A. Tohsaki,
\href{https://doi.org/10.1007/s11467-019-0917-0}{Front. Phys. 15 (2020) 14401}.

% Brink-Bloch wave function
\bibitem{Bri66}D.M. Brink, in {\it International School of Physics {\lq\lq}Enrico Fermi'', XXXVI} 
edited by C. Bloch (Academic Press, New York, 1966), p. 247.  

% DWIA framework, p,pn
\bibitem{Wak17}T. Wakasa, K. Ogata, T. Noro,
\href{https://doi.org/10.1016/j.ppnp.2017.06.002}{Prog. Part. Nucl. Phys. 96 (2017) 32-87}.
\bibitem{Phu19}N.T.T. Phuc, K. Yoshida, K. Ogata,
\href{https://doi.org/10.1103/PhysRevC.100.064604}{Phys. Rev. C 100 (2019) 064604}.
\bibitem{Aum13}T. Aumann, C.A. Bertulani, J. Ryckebusch,
\href{https://doi.org/10.1103/PhysRevC.88.064610}{Phys. Rev. C 88 (2013) 064610}.

% DWIA for recent experiments
\bibitem{Oli17}L. Olivier, S. Franchoo, M. Niikura {\it et al.},
\href{https://doi.org/10.1103/PhysRevLett.119.192501}{Phys. Rev. Lett. 119 (2017) 192501}.
\bibitem{Kaw18}S. Kawase, T. Uesaka, T.L. Tang {\it et al.},
\href{https://doi.org/10.1093/ptep/pty011}{Prog. Theor. Exp. Phys. 2018 (2018)}.
\bibitem{Che19}S. Chen, J. Lee, P. Doornenbal {\it et al.},
\href{https://doi.org/10.1103/PhysRevLett.123.142501}{Phys. Rev. Lett. 123 (2019) 142501}.
\bibitem{Tan19}R. Taniuchi, C. Santamaria, P. Doornenbal {\it et al.},
\href{https://doi.org/10.1038/s41586-019-1155-x}{Nature 569 (2019) 53–58}.
\bibitem{Tan20}T.L. Tang, T. Uesaka, S. Kawase {\it et al.},
\href{https://doi.org/10.1103/PhysRevLett.124.212502}{Phys. Rev. Lett. 124 (2020) 212502}.
\bibitem{Yan21}Z.H. Yang, Y. Kubota, A. Corsi {\it et al.},
\href{https://doi.org/10.1103/PhysRevLett.126.082501}{Phys. Rev. Lett. 126 (2021) 082501}.

% alpha knockout for Be isotopes
\bibitem{Lyu18}M. Lyu, K. Yoshida, Y. Kanada-En'yo, K. Ogata,
\href{https://doi.org/10.1103/PhysRevC.97.044612}{Phys. Rev. C 97 (2018) 044612}.
\bibitem{Lyu19}M. Lyu, K. Yoshida, Y. Kanada-En'yo, K. Ogata,
\href{https://doi.org/10.1103/PhysRevC.99.064610}{Phys. Rev. C 99 (2019) 064610}.

% alpha knockout, general
\bibitem{Yos16}K. Yoshida, K. Minomo, K. Ogata,
\href{https://doi.org/10.1103/PhysRevC.94.044604}{Phys. Rev. C 94 (2016) 044604}.
\bibitem{Yos18-2}K. Yoshida, K. Ogata, Y. Kanada-En'yo,
\href{https://doi.org/10.1103/PhysRevC.98.024614}{Phys. Rev. C 98 (2018) 024614}.
\bibitem{Yos19}K. Yoshida, Y. Chiba, M. Kimura, Y. Taniguchi, Y. Kanada-En'yo, K. Ogata,
\href{https://doi.org/10.1103/PhysRevC.100.044601}{Phys. Rev. C 100 (2019) 044601}.
\bibitem{Tan21}J. Tanaka, Z. Yang, S. Typel {\it et al.},
\href{https://doi.org/10.1126/science.abe4688}{Science 371 (2021) 260–264}.

\bibitem{Yos18}K. Yoshida, M. G\'{o}mez-Ramos, K. Ogata, A.M. Moro,
\href{https://doi.org/10.1103/PhysRevC.97.024608}{Phys. Rev. C 97 (2018) 024608}.

% definition of RWA
\bibitem{Hor77}H. Horiuchi, \href{https://doi.org/10.1143/PTPS.62.90}
{Prog. Theor. Phys. Supp. 62 (1977) 90-190.}

% Chiba-san's AMD model, N constrained
\bibitem{Chi15}Y. Chiba, M. Kimura,
\href{https://doi.org/10.1103/PhysRevC.91.061302}{Phys. Rev. C 91 (2015) 061302}.

% angular momentum projection
\bibitem{Hor86}H. Horiuchi, K. Ikeda, {Cluster model of the nucleus}, in:
T.T.S. Kuo, and E. Osnes (Eds.), Cluster Models and Other Topics, World
Scientific, Singapore, 1986, 1-258.

% alpha-alpha potential (Ali-Bodmer)
\bibitem{Ali66}S. Ali, A.R. Bodmer,
\href{https://doi.org/10.1016/0029-5582(66)90829-7}{Nucl. Phys. 80 (1966) 99-112}.

% laplace expansion method for RWA
\bibitem{Chi17}Y. Chiba, M. Kimura,
\href{https://doi.org/10.1093/ptep/ptx063}{Prog. Theor. Exp. Phys. 2017 (2017) 053D01}.

% new Brink expansion method for RWA
\bibitem{Lyu20}M. Lyu (unpublished).
% traditional Brink expansion method for RWA
\bibitem{Eny03b}Y. Kanada-En'yo, H. Horiuchi, 
\href{https://doi.org/10.1103/PhysRevC.68.014319}{Phys. Rev. C 68 (2003) 014319}.


% Franey-Love (p-n scattering)
\bibitem{FL85}M.A. Franey, W.G. Love,
\href{https://doi.org/10.1103/PhysRevC.31.488}{Phys. Rev. C 31 (1985) 488-498}.

% Dirac optical potential
\bibitem{Coo93}E.D. Cooper, S. Hama, B.C. Clark, R.L. Mercer,
\href{https://doi.org/10.1103/PhysRevC.47.297}{Phys. Rev. C 47 (1993) 297-311}.
% Single folding potential
\bibitem{Min10}K. Minomo, K. Ogata, M. Kohno, Y.R. Shimizu, M. Yahiro,
\href{https://doi.org/10.1088/0954-3899/37/8/085011}{J. Phys. G 37 (2010) 085011}.

\bibitem{Amos00}K. Amos, P.J. Dortmans, H.V. von Geramb, S. Karataglidis, J. Raynal,
\href{https://doi.org/10.1007/0-306-47101-9}{Adv. Nucl. Phys. 25 (2000) 276-536}.

\bibitem{Kel91}J.J. Kelly, A.E. Feldman, B.S. Flanders {\it et al.},
\href{https://doi.org/10.1103/PhysRevC.43.1272}{Phys. Rev. C 43 (1991) 1272-1287, (EXFOR C0142.006)}.

\bibitem{Coo09}E.D. Cooper, S. Hama, B.C. Clark,
\href{https://doi.org/10.1103/PhysRevC.80.034605}{Phys. Rev. C 80 (2009) 034605}.

% S.O effect in TDX
\bibitem{Nor19}T. Noro and K. Yoshida (private communication).
	
% Gogny interaction
\bibitem{Ber91}J. Berger, M. Girod, D. Gogny,
\href{https://doi.org/10.1016/0010-4655(91)90263-K.}{Comput. Phys. Commun. 63 (1991) 365-374}.

% Volkov interaction
\bibitem{Vol65}A.B. Volkov, 
\href{http://dx.doi.org/10.1016/0029-5582(65)90244-0}{Nucl. Phys. 74 (1965) 33-58}.

% G3RS interaction
\bibitem{Yam79}N. Yamaguchi, T. Kasahara, S. Nagata, Y. Akaishi, 
\href{http://dx.doi.org/10.1143/PTP.62.1018}{Prog. Theor. Phys. 62 (1979) 1018-1034}.

% 9Be experimental data
\bibitem{Til04}D.R. Tilley, J.H. Kelley, J.L. Godwin, D.J. Millener, J.E. Purcell, C.G. Sheu, H.R. Weller, 
\href{http://dx.doi.org/10.1016/j.nuclphysa.2004.09.059}{Nucl. Phys. 745 (2004) 155-362}.

\bibitem{Tsa05}M.B. Tsang, J. Lee, W.G. Lynch,
\href{https://doi.org/10.1103/PhysRevLett.95.222501}{Phys. Rev. Lett. 95 (2005) 222501}.

% Other models
\bibitem{Pie02}S.C. Pieper, K. Varga, R.B. Wiringa, 
\href{http://dx.doi.org/10.1103/PhysRevC.66.044310}{Phys. Rev. C 66 (2002) 044310}.
\bibitem{For05}C. Forss\'{e}n, P. Navr\'{a}til, W.E. Ormand, E. Caurier, 
\href{http://dx.doi.org/10.1103/PhysRevC.71.044312}{Phys. Rev. C 71 (2005) 044312}.
\bibitem{Del18}V. Della Rocca, F. Iachello, 
\href{http://dx.doi.org/10.1016/j.nuclphysa.2018.02.003}{Nucl. Phys. A 973 (2018) 1-32}.


\end{thebibliography}
\end{document}